\documentclass[12pt]{article}
\usepackage{amsmath}
\usepackage{amsfonts}
\usepackage{amssymb}
\usepackage{amsthm}
\usepackage{cite}
\usepackage{mathtools}
\usepackage{xcolor}
\usepackage{graphicx} 
\usepackage{caption}
\usepackage{cancel}

\usepackage[unicode=true,pdfusetitle,
 bookmarks=true,bookmarksnumbered=false,bookmarksopen=false,
 breaklinks=true,pdfborder={0 0 0},backref=false,colorlinks=true]
{hyperref}
\hypersetup{
 linkcolor=blue,citecolor=blue, urlcolor=blue}

\newcommand{\bcc}{\color{black}}
\newcommand{\bc}{\color{black} }
\newcommand{\oc}{\color{black} }

\newcommand{\mrm}{\mathrm}

\newlength{\dinwidth}
\newlength{\dinmargin}
\newcommand{\mcF}{\mathcal{F}}
\newcommand{\mcB}{\mathcal{B}}

\newcommand{\mcE}{\mathcal{E}}

\newcommand{\wh}{\widehat}
\newcommand{\pa}{\partial}

\newcommand{\vp}{\varphi}

\newcommand{\eps}{\varepsilon}

\newcommand{\nin}{\noindent}
\newcommand{\si}{\sigma}
\newcommand{\ph}{\phantom}

\newcommand{\fr}[2]{\frac{#1}{#2}}

\newcommand{\real}{\mathbb{R}}
\newcommand{\complex}{\mathbb{C}}
\newcommand{\la}{\lambda}
\newcommand{\non}{\nonumber}

\newcommand{\lan}{\langle}
\newcommand{\ran}{\rangle}

\def\proof{\noindent{\bf Proof. }}
\def\qed{$\Box$\medskip}

\newtheorem{theoreme}{Theorem } [section]
\newtheorem{proposition}[theoreme]{Proposition}
\newtheorem{lemma}[theoreme]{Lemma}
\newtheorem{definition}[theoreme]{Definition}
\newtheorem{corollary}[theoreme]{Corollary}
\newtheorem{remark}[theoreme]{Remark}
\newtheorem{example}[theoreme]{Example}
\newtheorem{criterion}[theoreme]{Criterion}

\newcommand{\beq}{\begin{equation}}
\newcommand{\eeq}{\end{equation}}
\newcommand{\beqa}{\begin{eqnarray}}
\newcommand{\eeqa}{\end{eqnarray}}
\newcommand{\ben}{\begin{arabicenumerate}}
\newcommand{\een}{\end{arabicenumerate}}
\newcommand{\bex}{\begin{example}}
\newcommand{\eex}{\end{example}}
\newcommand{\ber}{\begin{remark}}
\newcommand{\eer}{\end{remark}}
\newcommand{\bec}{\begin{corollary}}
\newcommand{\eec}{\end{corollary}}
\newcommand{\bep}{\begin{proposition}}
\newcommand{\eep}{\end{proposition}}
\newcommand{\becr}{\begin{criterion}}
\newcommand{\eecr}{\end{criterion}}

\def\bel{\begin{lemma}}
\def\eel{\end{lemma}}
\def\bet{\begin{theoreme}}
\def\eet{\end{theoreme}}
\def\bed{\begin{definition}}
\def\eed{\end{definition}}

\begin{document}
\title{A  soft-photon theorem for  the Maxwell-Lorentz system} 

\author{
{\bf Wojciech Dybalski}\\
Zentrum Mathematik, Technische Universit\"at M\"unchen,\\
E-mail: {\tt dybalski@ma.tum.de}
\and
{\bf  Duc Viet Hoang }\\
Fakult\"at f\"ur Physik, Ludwig-Maximilians-Universit\"at M\"unchen\\ 
E-mail: {\tt V.hoang@physik.lmu.de}}
\date{}

\maketitle
\newcommand{\F}{\mathfrak{F}}
\begin{abstract}
For the coupled system of classical Maxwell-Lorentz equations we show that the quantities
\beqa
{\bcc \F}(\hat x, t)=\lim_{|x|\to \infty} |x|^2 F(x,t), \quad  \mathcal{F}(\hat k, t)=\lim_{|k|\to 0} |k| \wh{F}(k,t), \non
\eeqa  
where  $F$ is the Faraday tensor, $\hat{F}$ its Fourier transform in space and $\hat{x}:=\fr{x}{|x|}$, are  independent of $t$. We combine this observation with the scattering theory for the Maxwell-Lorentz system due to Komech and Spohn,
which gives the asymptotic decoupling of $F$ into the scattered radiation  $F_{\mrm{sc},\pm}$ and the soliton field
$F_{v_{\pm\infty}}$ depending on the asymptotic velocity $v_{\pm\infty}$ of the electron  at large positive (+), resp. negative (-) times.
This gives a \emph{soft-photon theorem} of the form  
\beqa
\mathcal{F}_{\text{sc},+}(\hat{k}) - \mathcal{F}_{\text{sc},-}(\hat{k})= -(  \mathcal{F}_{v_{+\infty}}(\hat{k})-\mathcal{F}_{v_{-\infty}}(\hat{k})),\non
\eeqa
and analogously for $\F$, which links the low-frequency part of the scattered radiation to the change of the electron's velocity. 
 Implications for the infrared problem in QED
are discussed in the Conclusions.
\end{abstract}

\section{Introduction}
\setcounter{equation}{0}

It is well known that a formal application of the Noether theorem to the global  $U(1)$ symmetry
of QED gives conservation of the electric charge. It is less well known that a similar reasoning applied to the
local gauge symmetry ensures conservation of the spacelike asymptotic flux of the electric field\footnote{A {\bcc heuristic} argument can be found in https://en.wikipedia.org/wiki/Infraparticle. }
\beqa
\phi(n):=\lim_{r\to\infty} r^2n\cdot E(nr), \quad n\in S^2. \label{flux}
\eeqa
The relevance of such asymptotic quantities for qualitative understanding of  infrared problems
has been known for long \cite{Bu82}. Intriguing relations between asymptotic symmetries, 
soft-photon theorems and memory effects, recently pointed out  by  Strominger, led to a  revival of
interest in this subject (see \cite{St17} for a review). Thus there is every reason to advance rigorous
mathematical understanding of asymptotic constants of motion in classical and quantum electrodynamics.

In this paper we prove  the existence of a large family of asymptotic constants of motion, including (\ref{flux}), for a classical
system of coupled Maxwell-Lorentz equations, also known as the Abraham model (see Section~\ref{Preliminaries}). This system describes one (spatially extended) electron interacting via the Lorentz force with
the electromagnetic field.   The existence and uniqueness of solutions for this model and the long-time asymptotics
was clarified by the works of Komech and Spohn \cite{Sp, KS00} which provide the basis for the present investigation. These
authors emphasized the role of the traveling wave (or soliton) solutions, which have the form   
\beqa  
(E({\oc x,t}), B({\oc x,t}), q(t), v(t))= (E_{v}({\oc x}-q-vt),B_{v}({\oc x}-q-vt),q+vt,v),
\eeqa
where $E,B,q,v$ denote the electromagnetic fields, the position and velocity of the electron, respectively.
$E_v$ and $B_v$ are concrete functions which can be obtained by minimizing the total energy of the system at
fixed momentum. Of course, $E_{v=0}$ is simply the Coulomb field of the electron at rest. For arbitrary $|v|<1$ we have
(in Fourier space)
\begin{align}
\hat{E}_{v}(k) & = \frac{-ik+v(v\cdot ik)}{|k|^{2}-(k\cdot v)^{2}} e\hat{\varphi}(k),   \label{electric-field}
\end{align}
where $e\varphi\in \mathcal{C}_0^{\infty}(\real^3)$ describes the charge distribution of the electron. For sufficiently small $e$ and
 suitable initial data Komech and Spohn showed that
\beqa
\lim_{t\to\pm\infty}\|E(\,\cdot\,, t)-E_{v(t)}(\cdot-q(t))-E_{\mrm{sc}, {\bc \pm}}(\,\cdot\,,t)\|_{L^2(\real^3)}=0, \label{asymptotic-decoupling}
\eeqa
where the \emph{scattered radiation fields} $E_{\mrm{sc}, {\bc \pm}}$ satisfies the homogeneous Maxwell equations.  
The existence of the asymptotic velocities of the electron $v_{\pm\infty}=\lim_{t\to\pm\infty} v(t)$ is an
important intermediate result from \cite{Sp, KS00}.

 Let us now describe in more detail the findings of the present paper. Under slightly more restrictive assumptions than Komech and Spohn, we show that 
\beqa
\mathfrak{E}(\hat x, t)=\lim_{|x|\to \infty} |x|^2 E(x,t),  \label{space-constant-of motion}
\eeqa  
depends only on $\hat{x}:=x/|x|$ and is a constant of motion. That is, if the limit exist for $t=0$ then it exists for arbitrary $t\in \real$ and has the
same value. Clearly, the spacelike asymptotic flux of the electric field (\ref{flux}) inherits the properties of $\mathfrak{E}$.
For the discussion of infrared singularities it is convenient to have a counterpart of (\ref{space-constant-of motion}) in momentum space. We  introduce 
\beqa
\mathcal{E}(\hat k, t)=\lim_{|k|\to 0} |k| \wh{E}(k,t), \label{asymptotic-constants-of-motion}
\eeqa
which is a constant of motion in the same sense as $\mathfrak{E}$. 
We verify that it has a decomposition 
\beqa
\mathcal{E}(\hat{k},\pm t)  =  \mathcal{E}_{\text{sc}, \pm}(\hat{k}) +    \mathcal{E}_{v_{\pm\infty}}(\hat{k}), \quad t\geq 0,
\eeqa
which reflects the asymptotic decoupling in (\ref{asymptotic-decoupling}) for the two time directions. {\bcc Hence, due to the conservation of
$\mathcal{E}$ we obtain the following variant of the  \emph{soft-photon theorem} 
\beqa
\mathcal{E}_{\text{sc}, {\bc +}}(\hat{k}) - \mathcal{E}_{\text{sc},-}(\hat{k})= -(  \mathcal{E}_{v_{+\infty}}(\hat{k})-\mathcal{E}_{v_{-\infty}}(\hat{k}))
\label{soft-photon-theorem}
\eeqa
which links the low-frequency part of the scattered radiation to the change of the electron's velocity. Thus it has a similar
physical meaning {\oc as Weinberg's} soft-photon theorem in QED  (see e.g. \cite[formula (2.8.21)]{St17}) and there is currently substantial  
interest in the high energy physics community concerning  related asymptotic conditions in classical electrodynamics (see e.g. \cite{Pr18} and references therein).}
 While similar relations are {\bcc rigorously} known in the external current situation, see e.g. \cite{He17}, we are not aware of such results for the  Maxwell-Lorentz system. Since the r.h.s. of  (\ref{soft-photon-theorem}) can  readily be computed using (\ref{electric-field}), 
 we obtain  non-trivial information about the timelike asymptotics of  solutions of
the coupled Maxwell-Lorentz system:   Namely, at least one of the scattered fields  $E_{\mrm{sc}, \pm}$ (incoming or outgoing) must have a $1/|k|$ singularity for small $|k|$.  

{\bc  {\bcc Although} we consider only the Maxwell-Lorentz system with one electron in this paper, {\bcc we}  expect  that 
soft-photon theorems similar to (\ref{soft-photon-theorem}) are  true for a much larger class of systems. Natural  future
research directions include the case of many, possibly spinning electrons,  as discussed in \cite{Sp}. 
Actually, for any system of non-linear PDE, admitting long-time asymptotics in terms of soliton solutions, one can
try to formulate {\bcc and prove} relations similar to (\ref{soft-photon-theorem}). }

{\bc This paper is organized as follows. 
In Section~\ref{Preliminaries}  we provide some background material about the existence of solutions and scattering theory
of the  Maxwell-Lorentz system.
In Section~\ref{Conservation-space} we prove the existence of the constant of motion~(\ref{space-constant-of motion}) 
and in Section~\ref{Conservation-momentum} of its counterpart in momentum space (\ref{asymptotic-constants-of-motion}).
Section~\ref{Soft-photon-theorem} contains the proof of our main result, which is the soft-photon theorem~(\ref{soft-photon-theorem}).
In the Conclusions we discuss briefly  the infrared problem in QED from the perspective of our findings. 
}

\section{Preliminaries}\label{Preliminaries}
\setcounter{equation}{0}

We give an introduction to the Abraham model. All proofs can be found
in \cite{Sp, KS00, Ho}.

\subsection{The equations of motion for the {\bc Maxwell-Lorentz system}}

We are interested in the dynamics of well-localized charges representing
charged particles of finite extension. In the course {\bcc of this discussion}, let
$e$ be the \emph{charge} of {\oc such a} particle and $m$ its \emph{mass}.
{\bcc The \emph{position} $q:\mathbb{R} \to\mathbb{R}^{3}$ of the
particle is a function of time $t\in\mathbb{R}$ and 
 $\frac{d}{dt}q(t)\equiv\dot{q}(t)\eqqcolon v(t)$ denotes its
\emph{velocity}}. We always assume that $q\in\mathcal{C}^{2}(\mathbb{R})$
and $v\in\mathcal{C}^{1}(\mathbb{R})$, i.e., the second
derivative of  $t\mapsto q(t)$ and first derivative of $t\mapsto v(t)$ exist and are
continuous for all ${\bc t\in \real}$. The \emph{electric field} $E$ and the
\emph{magnetic field} $B$ are represented as vector fields $E=(E_{1},E_{2},E_{3})$
and $B=(B_{1},B_{2},B_{3})$ with 
\[
E_{i},B_{i}:\mathbb{R}^{3}\times\mathbb{R} \to\mathbb{R}\quad\text{for }i=1,2,3.
\]
Moreover, we {\bc introduce} $\varphi\in\mathcal{C}^{\infty}(\mathbb{R}^{3})$
{\bc which is the}  \emph{charge distribution} of the particle. {\bc That is}, for
all $(x,t)\in\mathbb{R}^{3}\times\mathbb{R}$
\begin{align*}
\rho(x,t) & =e\varphi\left(x-q(t)\right)\qquad\quad\quad\!\!\text{is the charge density,}\\
j(x,t) & =e\varphi\left(x-q(t)\right)v(t)\qquad\text{is the current density}.
\end{align*}
We assume in the Abraham model that the charge distribution 
$\varphi$ satisfies the following properties:
\begin{enumerate}
\item $\varphi$ is radial, i.e. $\varphi(x)=\varphi(|x|)$ for all $x\in\mathbb{R}^{3}$,
\item $\varphi$ is compactly supported, i.e. $\exists R_{\varphi}>0$ such
that for all $|x|\geq R_{\varphi} {\oc :}$ $\varphi{\bc \equiv}0$.  
\item $\varphi$ is {\bc normalized}, i.e. $\int \varphi(x)\,d^3x=1$. 
\end{enumerate}
\nin As a result, we are able to couple the Maxwell equations with the Lorentz
equation: 
\begin{definition}
We call 
\begin{align}
\partial_{t}B(x,t) & =-\nabla\times E(x,t),\\
\partial_{t}E(x,t) & =\nabla\times B(x,t)-e\varphi\left(x-q(t)\right)v(t),\\
\nabla\cdot E(x,t) & =e\varphi\left(x-q(t)\right),\\
\nabla\cdot B(x,t) & =0,\\
\frac{d}{dt}\left\{ m\gamma v(t)\right\}  & =e\left\{ E_{\varphi}\left(q(t),t\right)+v(t)\times B_{\varphi}\left(q(t),t\right) \right\} \label{Lorentz-force}
\end{align}
the \emph{equations of motion for the Abraham model}. Here $(x,t)\in\mathbb{R}^{3}\times\mathbb{R}$,
and $\gamma\equiv\gamma(t)\coloneqq\frac{1}{\sqrt{1-v(t)^2} }$. In addition, the fields $E$ and $B$ are smeared with the function
$\varphi$ in the last equation, {\bc that is},
\[
E_{\varphi}(x,t)\equiv\left(E(\cdot,t)\ast\varphi\right)(x),\qquad B_{\varphi}(x,t)\equiv\left(B(\cdot,t)\ast\varphi\right)(x).
\]
\end{definition}
\subsection{{\bc Existence of} solutions of the Maxwell-Lorentz system}

We will denote by $Y\coloneqq\left(E,B,q,v\right)$  a state of the Abraham model, and define the set of states
\beqa
\mathcal{L}\coloneqq\left\{ Y\in (L^{2}\cap \mathcal{C}^{2}) \times (L^{2} \cap \mathcal{C}^{2}  )\times \real^3\times\mathbb{V}\mid\|E\|+\|B\|+|q|+|\gamma v |<\infty\right\},\,\,\,
\eeqa
where $ \mathbb{V}\coloneqq\{v\in\mathbb{R}^{3}\mid|v|<1\}$, $\|F\|\coloneqq\left(\int d^3x\,|F(x)|^{2}\right)^{1/2}$ and $\mathcal{C}^2$ is the space of twice continuously differentiable functions. By imposing, in addition, the
constraints {\oc on the set of states}, we obtain the phase space of the model: 
\begin{align*}
\mathcal{M}\coloneqq\bigg\{ Y\in\mathcal{L}\mid 
 \nabla\cdot E(x)=e\varphi(x-q),\,\nabla\cdot B(x)=0\,\,\,\forall x\in\mathbb{R}^{3}\bigg\}.
\end{align*}
For future reference, we also introduce certain subsets of $\mathcal{M}$.
For any $\sigma\in[0,1]$  we say that $Y\in \mathcal{M}^{\si}$ if $Y\in \mathcal{M}$ 
and there exist $C,R>0$ such that for all $|x|>R $ 
\begin{align*}
 |E(x)|+|B(x)|+|x|\left(|\nabla E(x)|+|\nabla B(x)|\right)\leq\frac{C}{|x|^{1+\sigma}}.
\end{align*}
We can rewrite the equation of motion for the Abraham model as a generalized
differential equation
\[
\frac{d}{dt}Y(t)=F\left(Y(t)\right),\qquad Y(0)=Y^{0}
\]
or equivalently via integration {\bc over} the interval $[0,t]$, {\bc $t\in \real$}:
\[
Y(t)=Y^{0}+\int_{0}^{t}F\left(Y(s)\right)\,ds.
\]
Then we  have the following statement:
\begin{proposition}\label{existence}\cite{KS00, Sp}
Let $Y^{0}=(E^{0},B^{0},q^{0},v^{0})\in\mathcal{M}$. Then the integral
equation associated with the equation of motion, 
\[
Y(t)=Y^{0}+\int_{0}^{t}F\left(Y(s)\right)\,ds,
\]
has a unique solution $Y(t)=\left(E(\cdot,t),B(\cdot,t),q(t),v(t)\right)\in\mathcal{M}$
for all $t\in{\bc \mathbb{R}}$, which is continuous in $t$ and
satisfies $Y(0)=Y^{0}$.
\end{proposition}
{\bcc In \cite{KS00} the phase space $\mathcal{M}$ is defined without the restriction to  $\mathcal{C}^{2}$
functions. However, it easily follows from formulas (2.16), (2.17) of \cite{Sp}, that for initial conditions from $\mathcal{C}^{2}$
also the solutions are $\mathcal{C}^{2}$.}

{\bc 
 As the proof of Proposition~\ref{existence} from \cite{KS00, Sp} is restricted to the case $t\geq 0$, let us 
briefly discuss the general case. 
We recall that the proof of the proposition starts from the equation for the Lorentz force~(\ref{Lorentz-force}) 
and uses formulas  (2.16), (2.17) of \cite{Sp} to express the r.h.s. by integral formulas involving
propagators of the wave equation. For example, for $t\geq 0$ one expresses the electric field as
\begin{align}
E(t)&=\pa_t G_{\mrm{ret},t}* E^{0}+\nabla \times (G_{\mrm{ret},t}*B^{0})\label{initial-data-ret}\\
&\ph{4} -\int_0^t ds\,(\nabla G_{\mrm{ret},t-s}*\rho(s)+\pa_t G_{\mrm{ret},t-s}*j(s)), \label{current-ret}
\end{align}
where {\bcc we omitted the $x$ variable} and  $G_{\mrm{ret},t}(x):=\theta(t)\frac{1}{4\pi t}\delta(|x|-t)$ is the retarded propagator {\bcc of the wave equation}. For $t\leq 0$ this equation should be modified to
\begin{align}
E(t)&=\pa_t (-G_{\mrm{adv},t})* E^{0}+\nabla \times (-G_{\mrm{adv},t}*B^{0})\label{initial-data-adv}\\
&\ph{4} -\int_0^t ds\,(\nabla G_{\mrm{adv},t-s}*\rho(s)+\pa_t G_{\mrm{adv},t-s}*j(s)), \label{current-adv}
\end{align}
where $G_{\mrm{adv},t}(x):=G_{\mrm{ret},-t}(x)  =-\theta(-t)\frac{\delta(|x|+t)}{4\pi t}$ is the advanced propagator and   the minus sign in (\ref{initial-data-adv}), compared to (\ref{initial-data-ret}), is needed to match the
two formulas at $t=0$. (Of course, the parts (\ref{current-ret}) and (\ref{current-adv}) cannot have such a relative minus
sign, as they have to satisfy the same inhomogeneous wave equation). Thus a formula valid for all $t\in \real$ has the form
\begin{align}
E(t)&=\pa_t G_t* E^{0}+\nabla \times (G_{t}*B^{0})\label{initial-data-all}\\
&\ph{4} -\int_0^t ds\,(\nabla G_{ \mrm{ret}/\mrm{adv} ,t-s}*\rho(s)+\pa_t G_{ \mrm{ret}   / \mrm{adv},t-s}*j(s)), \label{current-all}
\end{align}
where  $G_t:=  G_{\mrm{ret},t}- G_{\mrm{adv},t}$ is the causal propagator and the choice of the retarded or advanced propagator
in (\ref{current-all}) is correlated with the sign of $t$.
The estimates involved in the later application of the Banach fixed point theorem in the proof of Proposition~\ref{existence} are insensitive 
to the changes of the propagator discussed above. Thus one obtains a unique trajectory $\real \ni t\mapsto q(t)$ and, via (\ref{initial-data-all})-(\ref{current-all}), the electric field. The magnetic field is treated analogously. Existence of the second derivative of $t\mapsto q(t)$ and the first derivative of 
$t\mapsto v(t)$ at $t=0$ follows  from the Lorentz force equation and the fact that the electromagnetic fields tend to $E^0,B^0$ as $t\to 0$ both
along the negative and positive values of $t$. {\bcc See also \cite{Ha18} for the problem of existence of solutions of the Maxwell-Lorentz system for
point charges for positive and negative times.}
}
\subsection{{\bc Soliton solutions}}
  {\bc Proposition~\ref{existence} gives the existence and uniqueness, but  no concrete construction of  solutions}.
To proceed, we consider a specific class of solutions, namely the
\emph{charge solitons} which represent charged particles with constant
velocities $v\in\mathbb{V}$. First, we define the following electromagnetic fields:
\begin{definition}\label{soliton-def}
We set for all $x\in\mathbb{R}^{3}\backslash\{0\}$
\[
\phi_{v}(x)\coloneqq\frac{e}{4\pi\sqrt{(x/\gamma)^{2}+(v\cdot x)^{2}}},\qquad\phi_{v\varphi}(x)\coloneqq (2\pi)^{-3/2} \phi_{v}\ast\varphi(x)
\]
and define the \emph{electric field }$E_{v}$\emph{ of a soliton with
velocity $v$} and the \emph{magnetic field} $B_{v}$ \emph{of a soliton
with velocity} $v\in \mathbb{V}$ as
\begin{align}
E_{v}(x)  =-\nabla\phi_{v\varphi}(x)+v\left(v\cdot\nabla\phi_{v\varphi}(x)\right), \quad B_{v}(x)  =-v\times\nabla\phi_{v\varphi}(x).
\end{align}
\end{definition}
The corresponding momentum space expressions are
\begin{align}
\hat{E}_{v}(k) =-ik\hat{\phi}_{v\varphi}+v(v\cdot ik\hat{\phi}_{v\varphi}), \quad   \hat{B}_{v}(k) =-v\times\left(ik\hat{\phi}_{v\varphi}\right),
\end{align}
where 
\begin{align}
\hat{\phi}_{v}\equiv\hat{\phi}_{v}(k)=\frac{e}{k^{2}-(k\cdot v)^{2}}, \quad
\hat{\phi}_{v\varphi} & \equiv\hat{\phi}_{v\varphi}(k)=  \hat{\phi}_{v}(k)\hat{\varphi}(k).
\end{align}
(Note that $\phi_{v}\not\in L^{1}(\mathbb{R}^{3},\mathbb{R})$, so
the Fourier  transformations\footnote{{\bcc We use the  conventions for the Fourier transform from \cite{RS2}.}} need to be understood in the distributional
sense). Now the soliton solution is constructed as follows:
\begin{proposition}\cite{KS00, Sp} For any $v\in \mathbb{V}$, the following family of states
\[
Y_{v}(t)=(E_{v}(\cdot-q-vt),B_{v}(\cdot-q-vt),q+vt,v)\in\mathcal{M}
\]
is a solution of the Abraham model. 
\end{proposition}

\subsection{Long-time asymptotics}


It turns out that a large class of solutions converges to  soliton solutions for large times {\bcc via emission of radiation}.
It is therefore convenient to define 
\begin{align} \label{Z-def}
\mathsf{Z}(x,t)=\left(\begin{array}{c}
Z_{1}(x,t)\\
Z_{2}(x,t)
\end{array}\right)=\left(\begin{array}{c}
E(x,t)-E_{v(t)}(x-q(t))\\
B(x,t)-B_{v(t)}(x-q(t))
\end{array}\right),
\end{align}
where $\left(E_{v(t)}(x-q(t)),B_{v(t)}(x-q(t)),q(t),v(t)\right)$
is the soliton approximation at time $t\in\mathbb{R}$. 
We remark for future reference that the system of equations  of the Abraham model can then be rewritten in the following form {\bc \cite{KS00, Sp}}\beqa
\mathsf{Z}(x,t)=\mathsf{U}(t)\mathsf{Z}(x,0)-\int_{0}^{t}\text{d}s\,\mathsf{U}(t-s)\mathsf{g}(x,s), \label{main-equation}
\eeqa
where
\[
\mathsf{g}(x,t)=\left(\begin{array}{c}
(\dot{v}(t)\cdot\nabla_{v})E_{{\bcc v(t)}}(x-q(t))\\
 (\dot{v}(t)\cdot\nabla_{v})B_{{\bcc v(t)}}(x-q(t))
\end{array}\right).
\]
The operator $\mathsf{U}$ is unitary on $L^{2}(\mathbb{R}^{3})\times L^{2}(\mathbb{R}^{3})$
and is given by the solutions of the homogeneous Maxwell's equations in position space:
\[
\mathsf{U}(t)\mathsf{F}(x, 0)\equiv\left(\begin{array}{c}
\left\{ \partial_{t}G_{t}\ast F_{1}(\cdot, 0)\right\} (x)+\nabla\times\left\{ G_{t}\ast F_{2}(\cdot,0)\right\} (x)\\
\left\{ \partial_{t}G_{t}\ast F_{2}(\cdot, 0)\right\} (x)-\nabla\times\left\{ G_{t}\ast F_{1}(\cdot, 0)\right\} (x)
\end{array}\right),
\]
{\bcc where $F_1, F_2$ are the components of  $\mathsf{F}$} and  $G_{t}(x)=\theta(t)\frac{1}{4\pi t}\delta(|x|-t) +{\bc \theta(-t)\frac{\delta(|x|+t)}{4\pi t}}$ is the {\bc causal} propagator. 
 We note for future reference that in momentum space
\beqa
\hat{G}_{t}(k)= (2\pi)^{-3/2} \fr{\sin {\oc (|k|t) }}{|k|}\quad  \textrm{ for $t\in \real$.} \label{the-causal-propagator}
\eeqa
The advantage of (\ref{main-equation})
is that the only implicit input on the r.h.s. is $t\mapsto (q(t),v(t))$. 

For large {\bc positive (+) or negative (-)} times, the differences on the r.h.s. of (\ref{Z-def}) should give the scattered radiation field. 
Thus we write $\mathsf{Z}_{\text{sc},{\bc \mrm{\pm}} }(x,t)=(E_{\text{sc},  {\bc \mrm{\pm}} }(x,t),B_{\text{sc}, {\bc \mrm{\pm}}}(x,t))$, where
\begin{align}
\mathsf{Z}_{\text{sc}, {\bc\pm}}(x,0)&:= \mathsf{Z}(x,0)-\int_{0}^{{\bc\pm}\infty}\text{d}s\,\mathsf{U}(-s)\mathsf{g}(x,s), \label{scattered-equation-zero}\\
\mathsf{Z}_{\text{sc}, {\bc\pm}  }(x,t)&:=\mathsf{U}(t)\mathsf{Z}_{\text{sc}, {\bc\pm}}(x,0).
\label{scattered-equation}
\end{align}
By definition, $(x,t)\mapsto \left(E_{\text{sc}, {\bc\pm} }(x,t),B_{\text{sc},  {\bc\pm}  }(x,t)\right)$ satisfy the homogeneous Maxwell equations.
The long-time asymptotics of the Abraham model is now described by the following theorem. {\bc We remark that the proof 
given in \cite{KS00, Sp} for $t\geq 0$ generalizes to $t\leq 0$, since the relevant estimates are not sensitive to the substitution 
$G_{\mrm{ret}}\to -G_{\mrm{adv}}$, cf. also formula~(\ref{the-causal-propagator}).}
\begin{theoreme}\cite{KS00, Sp} \label{velocity-theorem}
For   $|e|\leq\overline{e}$, where $\overline{e}$ is sufficiently small, and   $Y(0)\in\mathcal{M}^{\sigma}$
for $\sigma\in(0,1]$
the following statements hold true: 
\begin{enumerate}
\item The acceleration $\dot{v}$ satisfies $|\dot{v}(t)|\leq C(1+|t|)^{-1-\sigma}$ for some $C$ independent of $t$.
Hence, $\lim_{t\to{\bc\pm}\infty}v(t)=v_{{\bc\pm}\infty}\in  \mathbb{V}$ exists.
\item  For  $\left(E_{\mrm{sc}, {\bc\pm}},B_{\mrm{sc}, {\bc\pm}  }\right)$ defined in (\ref{scattered-equation-zero}), (\ref{scattered-equation}), we have
\begin{align}
\lim_{t\to{\bc \pm}\infty} & \bigg\{\|E(\,\cdot\,, t)-E_{v(t)}(\cdot-q(t))-E_{\mrm{sc},{\bc\pm}}(\,\cdot\,,t)\|\\
 & +\|B(\,\cdot\,,t)-B_{v(t)}(\cdot-q(t))-B_{\mrm{sc},\bc\pm}(\,\cdot\,, t)\|\bigg\}=0.
\end{align}
\end{enumerate}
{\bc (The $\pm$-signs above are correlated).}
\end{theoreme}
\nin\textbf{The assumptions of Theorem~\ref{velocity-theorem} are our standing assumptions in the remaining part of this
paper.}

\section{A conservation law in position space} \label{Conservation-space}
\setcounter{equation}{0}

{\bc We denote by $F$ the Faraday tensor, whose components are the electromagnetic fields $E_i$, $B_j$, $i,j=1,2,3$.
We set $\mathfrak{F}(x, t):=\lim_{|x|\to \infty} |x|^2 F(x,t)$ (if the limit exists) and denote by $\mathfrak{E}$, $\mathfrak{B}$ its electric and
magnetic components}.
\bet {\bc Suppose that $\F(x, 0):=\lim_{|x|\to \infty} |x|^2 F(x,0)$ exists and depends only on $\hat{x}=\fr{x}{|x|}$. Then also $\mathfrak{F}(\hat x, t):=\lim_{|x|\to \infty} |x|^2 F(x,t)$
exists for any $t\in \real$ and  depends only on $\hat{x}$ in the first variable. Moreover,  $\mathfrak{F}(\hat x,t)=\mathfrak{F}(\hat x,0)$.}
\eet
\begin{remark} In $\mathcal{M}^{\si=1}$ there are many examples of  initial data  {\bc with slow, Coulomb-type decay} for which $\mathfrak{F}(\hat x,0)$ exists.
\end{remark}
{\bcc \proof} {\bc  We treat only the electric field for $t>0$, as the remaining cases are analogous}. By (\ref{main-equation}),   we have
\beqa
E(x,t)=E_{1}(x,t)+E_{2}(x,t)+E_{3}(x,t)+E_{4}(x,t)+E_{{\bcc v(t)}}(x-q(t)) \label{main-equation-one}
\eeqa
with 
\begin{align*}
E_{1}(x,t) & :=\partial_{t}G_{t}\ast\left[E(\cdot,0)-E_{v(0)}(\cdot-q(0))\right](x),\\
E_{2}(x,t) & :=\nabla\times\left\{ G_{t}\ast\left[B(x,0)-B_{v(0)}(x-q(0))\right]\right\} ,\\
E_{3}(x,t) & :=-\int_{0}^{t}ds\left[\partial_{\tau}G_{\tau}\big|_{\tau=t-s}\ast(\dot{v}(s)\cdot\nabla_{v})E_{\bcc v(s)}(\cdot-q(s))(x)\right],\\
E_{4}(x,t) & :=-\int_{0}^{t}ds\left[\nabla\times\left\{ G_{\tau}\big|_{\tau=t-s}\ast(\dot{v}(s)\cdot\nabla_{v})B_{{\bcc v(s)}}(\cdot-q(s))(x)\right\} \right].
\end{align*}

Let us consider the contribution of $E_{1}$.   We 
obtain the chain of equalities below, which gives the existence of  $\lim_{|x|\to\infty}|x|^{2}E_{1}(x,t)$, using the existence of an 
analogous limit for the initial data and a soliton solution. We will use  in the first step  the limit $|x|\to \infty$ eliminates the $q(0)$ dependence.
We will also repetitively apply the  dominated convergence theorem to exchange the limit with integrals. Its assumptions are verified
using concrete formulas for the soliton fields and the fact that the initial data are in $\mathcal{M}^{\si}$, $\si\in (0,1]$. 
\begin{align*}
&\lim_{|x|\to\infty}|x|^{2}E_{1}(x,t) \\
& =-\lim_{|x|\to\infty}|x|^{2}\int d^{3}y\,\frac{\delta^{'}(|y|-t)}{4\pi|y|}\left[E(x-y,0)-E_{v_{0}}(x-y)\right]\\
 & =\lim_{|x|\to\infty}|x|^{2}\int d\Omega(\hat{y})\int d|y|\,\frac{\delta(|y|-t)}{4\pi}\frac{\text{d}}{\text{d}|y|}\left[|y|\left(E(x-y,0)-E_{v_{0}}(x-y)\right)\right]\\
 & =\lim_{|x|\to\infty}|x|^{2}\frac{1}{4\pi}\int d\Omega(\hat{y})\left[\left(E(x-\hat{y}t,0)-E_{v_{0}}(x-\hat{y}t)\right)+t\left(E'(x-\hat{y}t,0)-E_{v_{0}}'(x-\hat{y}t)\right)\right]\\
 & \stackrel{(\ast)}{=}\frac{1}{4\pi}\int d\Omega(\hat{y})\left[\lim_{|x|\to\infty}|x|^{2}E(x-\hat{y}t,0)-\lim_{|x|\to\infty}|x|^{2}E_{v_{0}}(x-\hat{y}t)\right]\\
 & =\frac{1}{4\pi}\int d\Omega(\hat{y})\left[\lim_{|x|\to\infty}|x|^{2}E\left(|x|\left(x-\frac{\hat{y}t}{|x|}\right),0\right)-\lim_{|x|\to\infty}|x|^{2}E_{v_{0}}\left(|x|\left(x-\frac{\hat{y}t}{|x|}\right)\right)\right]\\
 & =\lim_{|x|\to\infty}|x|^{2}E(x,0)-\lim_{|x|\to\infty}|x|^{2}E_{v_{0}}(x).
\end{align*}
The prime above denotes the derivative w.r.t. $|y|$ and in $(\ast)$ we used that $E'(x,0),E_{v_{0}}'(x)\sim\frac{1}{|x|^{2+\si}}$
for $\si>0$ as the initial data and the soliton solutions belong to $\mathcal{M}^{\si}$ for $\si\in (0,1]$. 

As for $E_2$, we find
\begin{align*}
\lim_{|x|\to\infty}|x|^{2}E_{2}(x,t) & =\lim_{|x|\to\infty}|x|^{2}\int d^{3}y\,\frac{\delta(|y|-t)}{4\pi|y|}\nabla\times\left[B(x-y,0)- B_{v_{0}}(x-y)\right]\\
 & =\lim_{|x|\to\infty}|x|^{2}\frac{1}{4\pi}\int d\Omega(\hat{y})\,t\nabla\times\left[B(x-\hat{y}t,0)-B_{v_{0}}(x-\hat{y}t)\right]\\
 & =\frac{1}{4\pi}\int d\Omega(\hat{y})\,t\,\lim_{|x|\to\infty}|x|^{2}\nabla\times\left[B(x-\hat{y}t,0)-B_{v_{0}}(x-\hat{y}t)\right]\\
 & =0,
\end{align*}
where we again made use of the fact that the initial data and the soliton solutions belong to $\mathcal{M}^{\si}$.

Let us move on to the contribution of $E_3$. First, we write
\[
\lim_{|x|\to\infty}|x|^{2}E_{3}(x,t)=-\lim_{|x|\to\infty}|x|^{2}\int_{0}^{t}ds\int d^{3}y\,\frac{(-1)}{4\pi|y|}\delta^{'}(|y|-(t-s))\left(\dot{v}(s)\cdot\nabla_{v}\right)E_{v(s)}(x-y),
\]
where  we used  that the limit (if it exists) eliminates the $q(s)$ dependence. Next, we note the equality
\begin{align*}
 & \int d^{3}y\,\frac{(-1)}{4\pi|y|}\delta^{'}(|y|-(t-s))\left(\dot{v}(s)\cdot\nabla_{v}\right)E_{v(s)}(x-y)\\
 & =\frac{1}{4\pi}\int d\Omega(\hat{y})\left[(t-s)\left(\dot{v}(s)\cdot\nabla_{v}\right)E_{v(s)}'(x-{\bcc \hat{y}}(t-s) )
 +\left(\dot{v}(s)\cdot\nabla_{v}\right)E_{v(s)}(x- {\bcc \hat{y} (t-s) } )\right].
\end{align*}
Hence, making use of the rapid decay of $E_{v(s)}'$, we find 
\begin{align*}
\lim_{|x|\to\infty}|x|^{2}E_{3}(x,t) & =\frac{-1}{4\pi}\int d\Omega(\hat{y}){\bcc \lim_{|x|\to\infty}
|x|^{2} } \int_{0}^{t}ds\,\left(\dot{v}(s)\cdot\nabla_{v}\right)
E_{v(s)}(x- {\bcc \hat{y} (t-s) })\\
 & =\frac{-1}{4\pi}\int d\Omega(\hat{y}){\bcc \lim_{|x|\to\infty}
|x|^{2} } \int_{0}^{t}ds\,\left(\dot{v}(s)\cdot\nabla_{v}\right)
E_{v(s)}(x)\\
 & =\frac{-1}{4\pi}\int d\Omega(\hat{y})\left[\lim_{|x|\to\infty}|x|^{2}E_{v(t)}(x)-\lim_{|x|\to\infty}|x|^{2}E_{v_{0}}(x)\right]\\
 & =-\lim_{|x|\to\infty}|x|^{2}E_{v(t)}(x)+\lim_{|x|\to\infty}|x|^{2}E_{v_{0}}(x),
\end{align*}
where we used the Fubini theorem in the first step, in the second step we used the presence of the limit to eliminate the shift by ${\bcc \hat{y} (t-s)}$
and in the third step we noted that the integral w.r.t. $s$ can be computed.  By reading the above computation backwards, we obtain the
existence of $\lim_{|x|\to\infty}|x|^{2}E_{3}(x,t)$.

As for $E_4$, we find 
\begin{align*}
&\lim_{|x|\to\infty}|x|^{2}E_{4}(x,t) \\
& = -\lim_{|x|\to\infty}|x|^{2}\fr{1}{{\bcc 4\pi}}\int_{0}^{t}ds\int d\Omega(\hat{y})\,(t-s)\left(\dot{v}(s)\cdot\nabla_{v}\right)\nabla\times B_{v(s)}(x-{\bcc \hat{y}(t-s)-q(s)})
  =0
\end{align*}
since $\nabla\times B_{v(t)}(x)\sim\frac{1}{|x|^{2+\si}}$ for
$\si>0$. 

By substituting  all the contributions above to (\ref{main-equation-one}), we obtain
\[
\lim_{|x|\to\infty}|x|^{2}E(x,t)=\lim_{|x|\to\infty}|x|^{2}E(x,0).
\]
This completes the proof. \qed

\section{A conservation law in momentum space} \label{Conservation-momentum}
\setcounter{equation}{0}

{\bc We recall that $F$ is the Faraday tensor
and we set $\mathcal{F}(k, t):=\lim_{|k|\to 0} |k| \hat{F}(k,t)$ (if the limit exists), where $\hat{F}$ is the Fourier transform of $F$ in space.
 We denote by $\mcE$, $\mcB$ the electric and magnetic components of $\mcF$}.
\bet\label{conservation-momentum} {\bc Suppose that $\mcF(k, 0):=\lim_{|k|\to 0} |k| \hat{F}(k,0)$ exists and depends only  on $\hat{k}=\fr{k}{|k|}$. 
Then also $\mcF(k, t):=\lim_{|k|\to \infty} |k| \hat{F}(k,t)$
exists for any $t\in \real$ and depends only on $\hat{k}$ in the first variable. Moreover, $\mcF(\hat k,t)=\mcF(\hat k,0)$.}
\eet
{\bcc \proof} {\bc Again, we consider only the electric field and the case $t>0$.} We recall formula (\ref{main-equation})
\begin{equation}
Z(x,t)=U(t)Z(x,0)-\int_{0}^{t}ds\,U(t-s)g(x,s).
\end{equation}
The electric part has the form considered already in the proof of the previous theorem:
\begin{align}
E(x,t) & =\partial_{t}G_{t}\ast\left[E(\cdot,0)-E_{v(0)}(\cdot-q(0))\right](x)\nonumber \\
 & \quad+\nabla\times\left\{ G_{t}\ast\left[B(x,0)-B_{v(0)}(x-q(0))\right]\right\} \nonumber \\
 & \quad-\int_{0}^{t}ds\bigg[\partial_{\tau}G_{\tau}\big|_{\tau=t-s}\ast(\dot{v}(s)\cdot\nabla_{v})E_{{\bcc v(s)}}(\cdot-q(s))(x)\nonumber \\
 & \qquad -\nabla\times\left\{ G_{\tau}\big|_{\tau=t-s}\ast(\dot{v}(s)\cdot\nabla_{v})B_{{\bcc v(s)}  }(\cdot-q(s))(x)\right\} \bigg]\nonumber \\
 & \quad+E_{{\bcc v(t)}}(x-q(t)). \non
\end{align}
Using (\ref{the-causal-propagator}), we obtain in momentum space
\begin{align}
\hat{E}(k,t) & =\cos(|k|t)\left[\hat{E}(k,0)-\hat{E}_{v(0)}(k)e^{{\bcc -}ikq(0)}\right]\nonumber \\
 & \quad+i\hat{k}\times\left\{ \sin(|k|t)\left[\hat{B}(k,0)-\hat{B}_{v(0)}(k)e^{{\bcc -i}kq(0)}\right]\right\} \nonumber \\
 & \quad-\int_{0}^{t}ds\bigg[\cos(|k|(t-s))\,(\dot{v}(s)\cdot\nabla_{v})\hat{E}_{{\bcc v(s)}}(k)e^{{\bcc -i}kq(s)}\nonumber \\
 & \qquad -i\hat{k}\times\left\{ \sin(|k|(t-s))\,(\dot{v}(s)\cdot\nabla_{v})\hat{B}_{{\bcc v(s)}}(k)e^{ {-\bcc i}kq(s)}\right\} \bigg]  \non\\
 & \quad+\hat{E}_{{\bcc v(t)}}(k)e^{{\bcc -}ikq({\bcc t})}.\non
\end{align}
Hence it holds
\begin{align}
\mathcal{E}(\hat{k},t) & =\mathcal{E}(\hat{k},0)-\mathcal{E}_{v_{0}}(\hat{k})+\mathcal{E}_{{\bcc v(t)}}(\hat{k})\nonumber \\
 & \quad-\lim_{|k|\to0}\int_{0}^{t}ds\bigg[\cos(|k|(t-s))\,(\dot{v}(s)\cdot\nabla_{v})|k|\hat{E}_{{\bcc v(s)}}(k)e^{{\bcc-i}kq(s)}\nonumber \\
 & \qquad - i{\bcc \hat{k}}\times\left\{ \sin(|k|(t-s))\,(\dot{v}(s)\cdot\nabla_{v}){\bcc |k|}\hat{B}_{{\bcc v(s)}}(k)e^{{\bcc-i}kq(s)}\right\} \bigg]\\
 & =\mathcal{E}(\hat{k},0)-\mathcal{E}_{v_{0}}(\hat{k})-\int_{0}^{t}ds\left[(\dot{v}(s)\cdot\nabla_{v})\mathcal{E}_{{\bcc v(s)}}(\hat{k})\right]+\mathcal{E}_{{\bcc v(t)}}(\hat{k})\non\\
 & =\mathcal{E}(\hat{k},0), \non
 \end{align}
 where we used in the second step $\lim_{|k|\to0} e^{{\bcc-i}kq(s)}=1$, $\lim_{|k|\to0} \cos(|k|(t-s))=1$, $\lim_{|k|\to0} \sin(|k|(t-s))=0$ and in the last step we
 noted that the integral w.r.t. $s$ can be evaluated. This concludes the proof. \qed
\section{{\bc Soft-photon theorem}} \label{Soft-photon-theorem}
\setcounter{equation}{0}

{\bc Now we are ready to state and prove our main result. We consider a solution $\real\ni t\mapsto Y(t)$ 
of the Maxwell-Lorentz system satisfying the assumptions of Theorem~\ref{velocity-theorem}. We denote by $F_{\mrm{sc},\pm}$
the scattered electromagnetic fields and by $F_{v_{\pm\infty}}$ the soliton solutions corresponding to the 
asymptotic velocities. }
\bet Under the assumptions of Theorem~\ref{conservation-momentum} the limits
\begin{align}
\mcF_{\mrm{sc},\pm}(k,t)=\lim_{|k|\to 0} |k| F_{\mrm{sc}}(k,t), \quad
\mcF_{ v_{\pm\infty} }(k,t)=\lim_{|k|\to 0} |k| F_{  v_{\pm\infty}  }(k,t) 
\end{align}
exist, depend only on $\hat k$ in the first variable and are independent of $t$. Moreover,
\beqa
\mathcal{F}_{\mrm{sc},+}(\hat{k}) +\mathcal{F}_{v_{+\infty}}(\hat{k})=\mathcal{F}_{\mrm{sc},-}(\hat{k})
+\mathcal{F}_{v_{-\infty} }(\hat{k}).
\eeqa
\eet
\begin{remark} An analogous statement holds  for $\F$. 
\end{remark}
\proof  {\bc We provide the details only for the case of the electric field}. For $t>0$,  the difference $Z(x,t)-Z_{\text{sc}}(x,t)$ yields
by (\ref{main-equation}), (\ref{scattered-equation}),
\begin{align}
E(x,t)-E_{\text{sc},+}(x,t) & =\int_{t}^{\infty}ds\bigg[\partial_{\tau}G_{\tau}\big|_{\tau=t-s}\ast(\dot{v}(s)\cdot\nabla_{v})E_{{\bcc v(s)}}(\cdot-q(s))(x)\nonumber \\
 & \qquad\nabla\times\left\{ G_{\tau}\big|_{\tau=t-s}\ast(\dot{v}(s)\cdot\nabla_{v})B_{{\bcc v(s)}}(\cdot-q(s))(x)\right\} \bigg]\non\\
 & +{ E_{v(t)}(x-q(t))}.
 \end{align}
Thus, we find by analogous arguments as in the proof of Theorem~\ref{conservation-momentum}, and making use of the fact that
$t\mapsto \dot{v}(t)$ is integrable (see Theorem~\ref{velocity-theorem}), that
\begin{equation}
\mathcal{E}(\hat{k})  -\mathcal{E}_{v(t)}(\hat{k})   -\mathcal{E}_{\text{sc},+}(\hat{k})=\mathcal{E}_{v_{+\infty}}(\hat{k})-\mathcal{E}_{v(t)}(\hat{k}),
\end{equation}
{\bc where all the taken limits exist and depend only on the indicated variables}. Consequently, 
\beqa
\mathcal{E}(\hat{k})  =  \mathcal{E}_{\text{sc},+}(\hat{k}) +    \mathcal{E}_{v_{+\infty}}(\hat{k})=  \mathcal{E}_{\text{sc},-}(\hat{k}) +  
  \mathcal{E}_{v_{-\infty}}(\hat{k}),  \label{positive-part-soft-photon-theorem}
\eeqa
where  in the last step we repeated the above reasoning for $t<0$ and made use of the fact that $\mathcal{E}(\hat{k})$ is a conserved quantity. \qed

\section{Conclusions}\label{Conclusions}
\setcounter{equation}{0}

In this paper we derived the following soft-photon theorem for the Maxwell-Lorentz system
\beqa
\mathcal{E}_{\text{sc},+}(\hat{k}) - \mathcal{E}_{\text{sc},-}(\hat{k})= -(  \mathcal{E}_{v_{+\infty}}(\hat{k})-\mathcal{E}_{v_{-\infty}}(\hat{k})),
\label{soft-photon-theorem-conc}
\eeqa
and analogously for the magnetic field. 
In order to elucidate its physical meaning, let us consider a scattering process in which the electron, which is initially at rest
($v_{-\infty}=0$), collides with infrared regular radiation ($\mathcal{E}_{\text{sc},-}(\hat{k})=0$). Making use of relation
(\ref{electric-field}) and considering that the l.h.s. of (\ref{soft-photon-theorem-conc}) is transverse, we obtain 
\beqa
\mathcal{E}_{\text{sc},+}(\hat{k})=-P_{\mrm{tr}}(\hat k)  \mathcal{E}_{v_{+\infty}}(\hat{k}), \label{transverse-formula}
\eeqa
where $P_{\mrm{tr}}(\hat k):=1-|\hat{k}\ran \lan \hat k|$ is the transverse projection.  This formula nicely brings together two 
seemingly distinct aspects of the infrared problem: the slow decay of the Coulomb field  on the r.h.s. and the infrared singularity of the 
scattered radiation on the l.h.s. It is clear from our discussion, that the presence of the asymptotic constants of motion (\ref{asymptotic-constants-of-motion}) is
behind this equality.
Making use of relation (\ref{electric-field}), we can rewrite (\ref{transverse-formula}) as follows 
\beqa
 \lim_{|k|\to 0} |k|\hat{E}_{\mrm{sc}, {\bcc +}}(k,t)= -\fr{ie}{  (2\pi)^{3/2}   } \bigg(   \fr{ (P_{\mrm{tr}}(\hat{k}) v_{\infty}) (\hat{k}\cdot v_{\infty})   }{1-( \hat{k}\cdot v_{\infty})^2 }    \bigg)
\label{classical-low-energy-behaviour}
 \eeqa
for any $t\geq 0$. This brings to light the infrared singularity $\hat{E}_{\mrm{sc}}(k,t)\sim 1/|k|$ of the scattered radiation. 
By an analogous discussion for the magnetic field we get
\beqa
 \lim_{|k|\to 0} |k|\hat{B}_{\mrm{sc}, {\bcc +}}(k,t)=\fr{e}{ (2\pi)^{3/2}} \fr{v_{\infty}\times i\hat{k}}{1- (\hat k\cdot v_{\infty})^2}, \label{classical-B}
\eeqa
thus it also exhibits a $1/|k|$ behavior for small momenta.

It is customary to say that such radiation fields `escape from the Fock space'  of the quantised variant of our theory.
Although such infrared jargon may be clear for experts, let us take the opportunity here to explain it from the perspective of
the present investigation.
Consider the  free, second-quantised  electromagnetic  fields (in Fourier space), which are operator valued distributions on Fock space 
\beqa
& &\hat{ \mathsf{E}}(k,t):=\sum_{\la=\pm} \sqrt{ \fr{|k|}{2} } (i\eps_{\la}(k) e^{-i|k|t} a_{\la}(k)- i\eps_{\la}(-k) e^{i|k|t}a_{\la}^*(-k)), \label{quantum-electric}\\
& &\hat{\mathsf{B}}(k,t):= \sum_{\la=\pm} \sqrt{\fr{1}{2|k|}} ik\times \big(\eps_{\la}(k)e^{-i|k|t}a_{\la}(k) +\eps_{\la}(-k) e^{i|k|t} a_{\la}^*(-k)\big). \label{quantum-magnetic}
\eeqa
 Here $a_{\la}^{(*)}(k)$ are the creation and annihilation operators of photons with momentum $k$ and polarisation $\la$,
$\eps_{\la}(k)$ are the polarisation vectors and we refer e.g. to \cite{Sp} for more details {\bcc on these formulas}.  We interpret these fields as the (outgoing) asymptotic fields of the theory. 
To study the classical limit of this quantum theory, we consider 
expectation values of (\ref{quantum-electric}),   (\ref{quantum-magnetic}) on coherent states of the form
\beqa
|w\ran=\exp\bigg({\sum_{\la=\pm}\int d^3k \big(w(k)\cdot \eps_{\la}(k) a_{\la}^*(k)  - \mrm{h.c.}  \big)   }\bigg)|0\ran, \label{coherent-state}
\eeqa
where $|0\ran$ is the Fock space vacuum. ({\bc For an extensive discussion of the role of coherent states in the study of infrared problems we
refer to \cite{FMS79}}). Specifically, we are looking for such $\complex^3$-valued functions $w$ that this expectation
value reproduces the classical low-energy behaviour (\ref{classical-low-energy-behaviour}), (\ref{classical-B})
\beqa
& &\lim_{k\to 0}|k|\lan w |\hat{ \mathsf{E}}(k,t)| w\ran= \lim_{|k|\to 0} |k|\hat{E}_{\mrm{sc},  {\bcc +} }(k,t),  \label{coherent-states-one} \\
 & & \lim_{|k|\to 0}|k|\lan w |\hat{ \mathsf{B}}(k,t)| w\ran= \lim_{|k|\to 0} |k|\hat{B}_{\mrm{sc}, {\bcc +} }(k,t),
\label{coherent-states-two}
\eeqa
for $t\geq 0$. The l.h.s.  of  (\ref{coherent-states-one}), (\ref{coherent-states-two}) are readily computed and the equality holds for $w$ of the familiar form, {\bcc 
see e.g. }\cite{Fr73,Fr74.1, CFP09, CFP07, FP10,KM13},
\beqa
w(k)=- \fr{e\hat\vp(k)   }{\sqrt{2}|k|^{3/2}} \fr{v_{\infty}}{1-\hat{k}\cdot v_{\infty}}. \label{w}
\eeqa
As the $|k|^{-3/2}$ singularity in (\ref{w}) is not square integrable, indeed $|w\ran$ given by (\ref{coherent-state}) is not 
a  vector in Fock space, but only makes sense as a state $\lan w|\,\cdot\,| w\ran$ on the algebra of the electromagnetic fields. 
Via the GNS construction it gives a representation of this algebra which is disjoint from the Fock representation. In this
sense the scattered field escapes from the Fock space.

Since $E_{\mrm{sc}, {\bcc \pm}}$ appear in (\ref{asymptotic-decoupling}) as  long-time asymptotics 
of the solutions of the Maxwell-Lorentz system, one can ask if $|w\ran$ can be obtained by scattering
theory. This turns out to be the case, as we have the equality (up to a phase), {\bcc which can be traced  back to  Bloch and Nordsieck} \cite{BN37,BN37.1}
\beqa
|w\ran=W^+|0\ran, \quad W^+=\mrm{Texp}\big(i\int_0^{\infty}dt\int d^3x \, \mathsf{A}(x,t) \cdot j_{v_{\infty}}(x,t)  \big). \label{Dyson}
\eeqa
{\bcc Here} $\mathsf{A}$ is the electromagnetic potential in the Coulomb gauge and $j_{v_{\infty}}(x,t):=ev_{\infty} \vp(x- v_{\infty}t)$ is the
external current describing the electron moving with velocity $v_{\infty}$. We recognize that $W^+$ is (up to a phase) the outgoing 
Dyson wave operator of  the second-{\oc quantised} electromagnetic field interacting with such a current. Here the infrared singularity
is hidden in a possible divergence of the time-integral and the equality in (\ref{Dyson}) is meant in the sense of
states on the algebra of the electromagnetic fields. Formula (\ref{Dyson}) is at the basis of the Faddeev-Kulish approach \cite{FK70, Fr73, Dy17},
as it ensures that the Dollard modifier maps the Fock space vacuum to the right representation of  this algebra. A variant of this formula was also
used in \cite{CD19} to show that this representation cannot be distinguished from the vacuum upon restriction to the interior of
a future or backward lightcone. {\bc This latter observation} supports the Buchholz-Roberts approach {\bcc to infrared problems} \cite{BR14}.

\vspace{0.5cm}

{\bc \nin\textbf{Acknowledgements:} W.D. would like to thank Herbert Spohn for useful comments.
Thanks are also due to Annika Steibel for helpful discussions on the existence and uniqueness of solutions
of the Maxwell-Lorentz system. This work was  partially supported by the Deutsche Forschungsgemeinchaft (DFG) 
within the Emmy Noether grant DY107/2-1.}

\end{document}